# Electronic structure of BaSnO$_3$ investigated by high-energy-resolution electron energy-loss spectroscopy and *ab initio* calculations

**Authors:** Hwanhui Yun, Mehmet Topsakal, Abhinav Prakash, Koustav Ganguly, Chris Leighton, Bharat Jalan, Renata M. Wentzcovitch, K. Andre Mkhoyan*, and Jong Seok Jeong*

Department of Chemical Engineering and Materials Science, University of Minnesota, Minneapolis, Minnesota 55455, USA

*Corresponding Authors: mkhoyan@umn.edu (K.A.M), jsjeong@umn.edu (J.S.J)

**ABSTRACT**

There has been growing interest in perovskite BaSnO$_3$ due to its desirable properties for oxide electronic devices including high electron mobility at room temperature and optical transparency. As these electronic and optical properties originate largely from the electronic structure of the material, here the basic electronic structure of epitaxially-grown BaSnO$_3$ films is studied using high-energy-resolution electron energy-loss spectroscopy in a transmission electron microscope and *ab initio* calculations. This study provides a detailed description of the dielectric function of BaSnO$_3$, including the energies of bulk plasmon excitations and critical interband electronic transitions, the band structure and partial densities of states, the measured band gap, and more. To make the study representative of a variety of deposition methods, results from BaSnO$_3$ films grown by both hybrid molecular beam epitaxy and high pressure oxygen sputter deposition are reported.

# I. INTRODUCTION

Perovskite structure BaSnO$_3$ has gained increasing attention recently as a candidate material for next-generation oxide electronic devices. This material has shown notably high electron mobility at room temperature, reported to be up to 320 cm$^2$V$^{-1}$s$^{-1}$ in bulk single crystal [1,2] and up to 150 cm$^2$V$^{-1}$s$^{-1}$ in epitaxial thin films of La-doped BaSnO$_3$ [3], with good optical transparency in the visible region. These properties are desirable in transparent conducting materials for solar cells, displays, high-mobility channels for transistors, and other applications [4-8]. Recent successes in thin film growth of epitaxial BaSnO$_3$ using vacuum deposition techniques [3,9-11] have opened up new possibilities to engineer various heterostructures based on this material, and to improve its electron mobility through defect control, polarization doping, etc. [6,8,12,13]. However, to optimize the potential utility of this material, it is essential to understand the electronic structure of BaSnO$_3$ in order to elucidate the mechanisms for high mobility. To this end, there have been several previous *ab initio* studies and optical measurements to understand various properties of BaSnO$_3$. Electronic band structures have been calculated [14-17], and the conduction band edge effective mass [18], band gap energy [8,19,20], and dielectric constant [21] have been experimentally determined.

Here, we report comprehensive information on the electronic structure of BaSnO$_3$ using high-energy-resolution electron energy-loss spectroscopy (EELS) in a scanning transmission electron microscope (STEM). We have acquired high-energy-resolution EELS spectra at elemental edges and compared them with *ab initio* predictions to examine the electronic structure of BaSnO$_3$. The low-loss region (0-50 eV) of EELS has been used to study dielectric properties, including interband electronic transitions and plasmon excitations. In the high-loss region (>50 eV), fine structure of core-edges reveals the details of the conduction band by directly probing the element-specific empty density of states (DOS) of the material. Good agreement between experimentally measured EELS spectra and the results of *ab* *initio* calculations provides critical reliability for the details of the electronic structure of BaSnO$_3$ reported here.

## II. METHODS

**A. Sample preparation**

Two epitaxial, single-phase BaSnO$_3$ films were grown on SrTiO$_3$(001) and LaAlO$_3$(001) substrates by hybrid molecular beam epitaxy (MBE) [9] and high pressure oxygen sputter deposition [22], respectively. A nominal thickness was 72 and 28 nm for the BaSnO$_3$ films on SrTiO$_3$ and LaAlO$_3$, respectively. The as-grown BaSnO$_3$ film on LaAlO$_3$ was further annealed in ultra-high vacuum (<10$^7$ Torr) at 900 °C for 4 h. This process dopes the BaSnO$_3$ film with oxygen vacancy to ~5×10$^{19}$cm$^{-3}$ [22]. Cross-sectional transmission electron microscopy samples were prepared by using a focused ion beam (FEI Quanta 200 3D) lift out method, where the samples were thinned by a 30 kV Ga-ion beam and then cleaned with a 5 kV Ga-ion beam. The samples were further polished by Ar-ion milling using a Fischione 1010 ion mill and a Gatan precision ion polishing system. This process of sample preparation provides relatively damage-free, electron-transparent samples with thickness in the range 20-60 nm. The relative sample thickness (t) was estimated to be t/$\lambda_P$ = 0.25-0.50 [23] using the EELS log-ratio method ($\lambda_P$ is mean-free-path of plasmon excitations) [24].

**B. STEM imaging and EELS data acquisition**

STEM imaging and EELS measurements were carried out using an aberration-corrected FEI Titan G2 60-300 (S)TEM equipped with a CEOS DCOR probe corrector, a Schottky extreme field emission gun (X-FEG), a monochromator, and a Gatan Enfinium ER spectrometer. The microscope was operated at 200 keV. The semi-convergence angle of the STEM probe was 15 mrad and the beam current was set to 50 pA. High-angle annular dark-field (HAADF) images were recorded with a detector angle of 42-200 mrad; the collection angle of the EELS was 29 mrad. A dual EELS mode, which simultaneously collects the low-loss region including a zero-loss peak (ZLP) and high-loss region, was used to correct the energy alignment, when needed. Energy dispersions of 0.05 and 0.1 eV per channel were used to measure low-loss and core-loss data of both $BaSnO_3$ films, while EELS data for detailed analysis of interband electronic transitions were acquired with an energy dispersion of 0.01 eV per channel. Energy resolutions for these dispersion values, estimated from the full-width at half maximum (FWHM) of the ZLPs, were 0.4, 0.25, and 0.13 eV for 0.1, 0.05, and 0.01 eV per channel, respectively. For an accurate determination of peak positions, the alignment and dispersion values of the spectrometer were calibrated using reference peaks of Si $L_{2,3}$ of $SiO_2$ (108.3 eV), $\pi^*$ of graphite (285.37 eV), and Ni $L_3$ of NiO (852.75 eV) [24]. Several EELS spectra from equivalent regions were acquired and summed to reduce the effects from very small, but unavoidable specimen damage by the STEM electron beam and to increase the signal-to-noise ratio. All EELS spectra were obtained from the central regions of $BaSnO_3$ films in order to minimize the influence of the film-substrate interface and surface.

**C. *Ab initio* calculations**

*Ab initio* calculations were performed using the WIEN2K code [25-27]. A generalized gradient approximation (GGA) using Perdew-Burke-Ernzenhof (PBE) parametrization [28] was adopted for the electronic exchange and correlation functional. The Brillouin zone was sampled at a 16×16×16 shifted *k*-point grid using the tetrahedron method [29]. The wave functions were expanded in spherical harmonics inside non-overlapping atomic spheres of radius $R_{MT}$ (muffin-tin radii) and in plane waves for the remaining space of the unit cell. $R_{MT}$ values for Ba, Sn, and O were set at WIEN2K defaults. The plane wave expansion in the interstitial region was determined by a cut-off wave vector chosen to be $k_{max} = 7.0/R_{MT}$. Empty states up to 4.0 Ry (= 54.4 eV) above the Fermi level, $E_F$, were included in the calculations. Frequency dependent dielectric matrix calculations within the independent particle random phase approximation were performed using the OPTIC module of WIEN2K, as detailed in Ref. [30]. O $K$ edge EELS simulations were performed using the TELNES3 module of WIEN2K with calculation parameters from the experimental setup. Core-hole effects were included. Because calculations using GGA are known to result in a smaller band gap than experimental results [15,17], the band gap was adjusted to the measured value.

## III. RESULTS AND DISCUSSION

Figure 1 shows HAADF-STEM images of the $BaSnO_3$ films that were used for EELS analysis in this work. From the HAADF-STEM images, the film thickness was measured to be 72 and 30 nm for the $BaSnO_3$ films on $SrTiO_3$ and $LaAlO_3$, respectively. High-resolution HAADF-STEM images demonstrate that these $BaSnO_3$ films are indeed epitaxially grown on the substrates, and form a sharp interface with relatively low surface roughness.

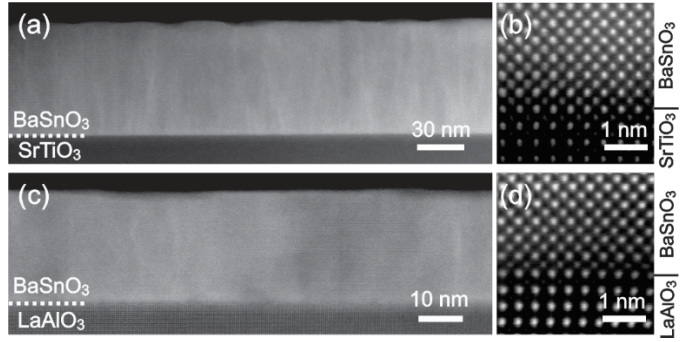

FIG. 1. Low-magnification and high-resolution HAADF-STEM images of BaSnO$_3$ films: (a,b) 72 nm-thick BaSnO$_3$ film grown on a SrTiO$_3$(001) substrate by hybrid MBE and (c,d) 30 nm-thick BaSnO$_3$ film grown on a LaAlO$_3$(001) substrate by high pressure oxygen sputter deposition. Atomic-resolution HAADF-STEM images of each film at the interfaces are shown in (b) and (d). The images were filtered using a standard Wiener filter, where a Butterworth filter was applied to remove high-frequency noise and amorphous components of the images [31,32]. The parameters used for the filter were: 5 for the step size of the iteration; 5 % for delta % (the upper limit of the % of pixels to be changed by each iteration); 20 for the number of cycles; 2 for the order of the Butterworth filter; and 0.5 for the FWHM of the Butterworth filter.

## A. Calculated dielectric function and band structure

The calculated dielectric function of BaSnO$_3$ is shown in Fig. 2. The complex dielectric function $\varepsilon(E) = \varepsilon_1(E) + i\varepsilon_2(E)$ describes how the electron gas in materials responds to an applied electromagnetic field [24,33-38]. The energy where the real part of the dielectric function, $\varepsilon_1(E)$, changes from negative to positive, evidences a bulk plasmon due to collective oscillations of the electrons. The calculated $\varepsilon_1(E)$ indicates that in BaSnO$_3$ bulk plasmons are generated at 15.2 and 26.6 eV, marked as I and E$_P$ in Fig. 2, respectively. In the imaginary part of the dielectric function, $\varepsilon_2(E)$, peaks correspond to the direct interband electronic transitions [24,33-38]. Distinguishable peaks in $\varepsilon_2(E)$ are marked with arrows and labeled in Fig. 2, and details are discussed in later sections.

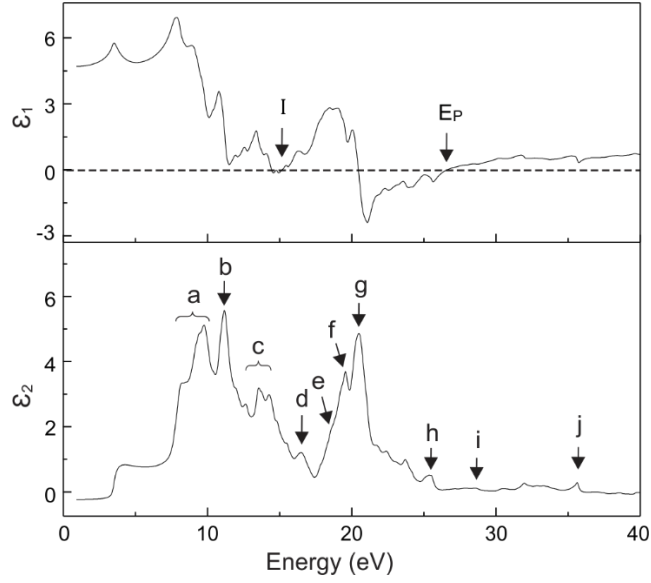

FIG. 2. The calculated real, $\varepsilon_1$, and imaginary, $\varepsilon_2$, parts of the dielectric function of BaSnO$_3$. The bulk plasmon energies (I, E$_P$) in $\varepsilon_1$ and direct interband transition peaks (a to j) in $\varepsilon_2$ are marked.

The calculated band structure and DOS of BaSnO$_3$ are shown in Fig. 3. The results are essentially consistent with previous reports [16,17,19,39,40]. The conduction band minimum is positioned at the Γ point (0, 0, 0) and the valence band maximum is at the M point (0.5, 0.5, 0) in reciprocal space. Figure 3(b) shows that the lower conduction bands from 3 to 8 eV consist predominantly of *s* and *p* states of Sn and O, and the conduction bands above 8 eV are mainly composed of *d* and *f* states of Ba. The O 2*p* states are dominant in the upper valence band, and high-density Ba 5*p* states are found at around −10 eV.

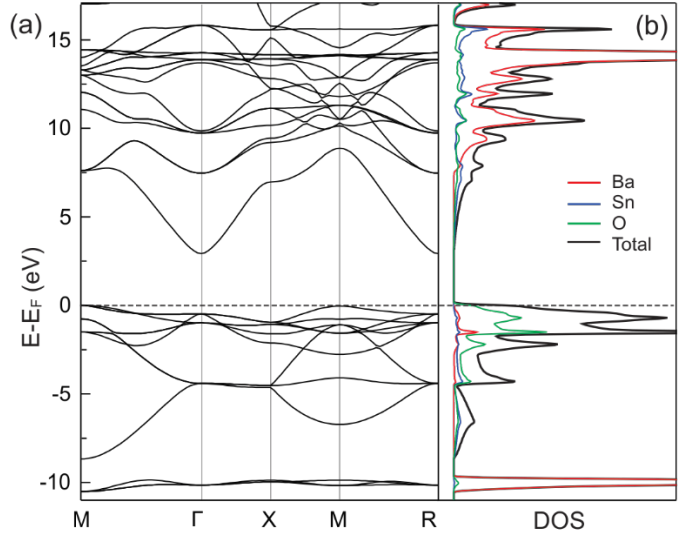

FIG. 3. Calculated band structure (a) and elementally resolved and total DOS (b) of $BaSnO_3$. Here $E_F$ is at the valence band maximum.

**B. Band gap measurement**

The band gap of $BaSnO_3$ films was measured using high-energy-resolution low-loss EELS data. To measure the band gap from the low-loss EELS data, the ZLP was removed as follows: (1) a zero-loss EELS spectrum was acquired in vacuum, (2) it was scaled to fit with the low-loss EELS spectra in the 1 to 2.5 eV range using a multiple linear regression method, and (3) the fitted zero-loss EELS spectrum was subtracted from the low-loss EELS data. To improve statistics, 22 spectra from the $BaSnO_3$ film on $SrTiO_3$ and 14 spectra from the $BaSnO_3$ film on $LaAlO_3$ were acquired and analyzed. One representative example is shown in Fig. 4. The onset value of the ZLP-subtracted spectrum indicates the band gap energy. The measured band gap energy for $BaSnO_3$ films on $SrTiO_3$ and $LaAlO_3$ substrates was $3.0 \pm 0.1$ eV and $3.1 \pm 0.2$ eV, respectively. To confirm the validity of this method, the same analysis was applied to the $SrTiO_3$ substrate and the resulting band gap energy was $3.3 \pm 0.1$ eV, which is consistent with well documented $SrTiO_3$ band gap energy of 3.25 eV [41]. The band gap energy of $BaSnO_3$ films from our measurements are compared with the values reported in the literature in Table I. A band gap energy of 3.0 eV is used for all subsequent analyses.

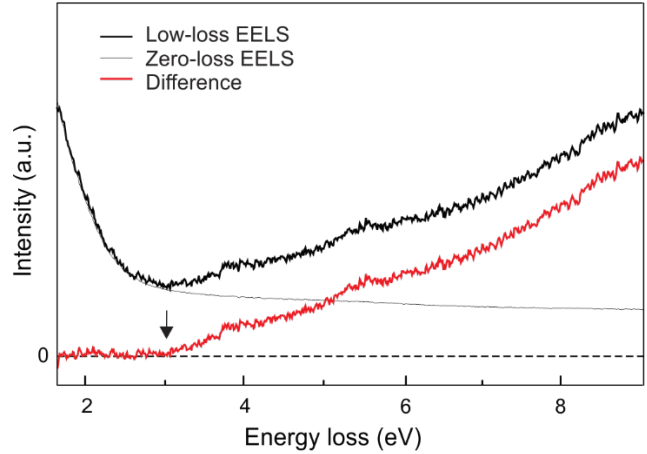

FIG. 4. A high-energy-resolution low-loss EELS spectrum obtained from a BaSnO$_3$ film on a LaAlO$_3$ substrate. The zero-loss EELS, obtained in vacuum, was fitted to the low-loss EELS spectrum and then subtracted from the spectrum to get the difference. The onset of the difference, which corresponds to the band gap, is indicated by an arrow.

TABLE I. Experimentally-measured band gap energy ($E_g$) of BaSnO$_3$ from the literature.

| $E_g$ (eV) | | Method | Ref. |
| --- | --- | --- | --- |
| Direct | Indirect | | |
|  | 3.0 ± 0.1 | Low-loss EELS | this work |
|  | 3.1 ± 0.2 | | |
| 3.56 ± 0.05 | 2.93 ± 0.05 | Ellipsometry | [8] |
| 3.12 | 2.85 | Diffuse reflectance | [42] |
| 3.1 | 2.95 | Transmittance | [43] |
| 3.4 |  | Reflectance | [38] |

**C. Low-loss EELS analysis**

Low-loss EELS spectra were measured on the BaSnO$_3$ films and compared with the electron energy-loss function calculated from the dielectric function. The energy-loss spectrum of incident probe electrons, which travel through a sample and act as an electromagnetic wave, can be deduced from the imaginary part of the inverse dielectric function, as it is proportional to $Im[-1/\varepsilon]$ (or $\varepsilon_2/(\varepsilon_1^2 + \varepsilon_2^2)$) [24,33-38]. To account for the energy spread of the incident electron beam, the calculated $Im[-1/\varepsilon]$ was convolved with a Gaussian function [44]. The FWHM of the Gaussian function was set to be the FWHM of the ZLP for each data set. The measured low-loss EELS spectra from $BaSnO_3$ films and the calculated $Im[-1/\varepsilon]$ are then compared, as can be seen in Fig. 5. For more quantitative analysis, the peak positions are compared in Table II. The overall shape and peak positions from the calculation and the experimental measurements are in good agreement. However, there are noticeable differences between the calculated $Im[-1/\varepsilon]$ and the measured low-loss EELS spectra because: (1) the calculation underestimates damping of plasmon oscillations [24,33] and (2) contributions from surface plasmons, Cerenkov radiation [45,46], and core-loss Sn $N_{4,5}$ edge are not included in the calculation.

In low-loss EELS, a bulk plasmon peak is observed at 26.1 eV (denoted as $E_P$ in Fig. 5), whereas it appears at 26.6 eV in the calculated $Im[-1/\varepsilon]$ (see Fig. 2). This discrepancy is due to plasmon damping. Plasmon damping induces broadening and shift of plasmon peaks; the plasmon linewidth, $\Delta E_P$, which is the FWHM of a plasmon peak, is determined by the plasmon damping constant, and the plasmon peak position shifts from $E_P$ to a lower value of $\sqrt{E_P^2 - (\Delta E_P^2/2)}$ [24]. The plasmon damping constant is governed by how fast plasmons decay by transferring their energy to an alternative mechanism, in the majority of cases to interband electronic excitations [24]. In a film, the energy transfer to the interband electronic excitations is facilitated by the extra energy or momentum from the lattice or phonons, which is followed by the faster plasmon damping. In addition, plasmons also decay via defects present in the films and through direct generation of phonons [47,48]. Because phonons and defects are not considered in the calculations, the experimental plasmon peak appears at the lower energy of 26.1 eV compared to the calculated value of 26.6 eV [49]. An additional plasmon peak at 15.2 eV is also predicted from the calculated dielectric function, as discussed above. We speculate that this plasmon represents oscillations of electrons in a subsidiary band near the valence band, but this needs further study. Due to several peaks from interband transitions and surface plasmons, this plasmon peak is not clearly identifiable in the low-loss EELS data. As stated above, the experimental low-loss EELS includes other contributions, such as surface plasmon peaks and the Sn $N_{4,5}$ edge, in contrast to the calculation. The surface plasmon peak is expected to be at around 18.8 eV from $E_{SP} = E_P/\sqrt{2}$ [24,33,34,38]; however, such a peak is not discernible as it is superimposed on strong interband transition peaks. The Sn $N_{4,5}$ edge should also be present starting from around 24 eV [50], but is not readily apparent due to overlap with the bulk plasmon and interband transition peaks.

The peaks, labeled from a to k in low-loss EELS in Fig. 5(a), result from interband electronic transitions (see Fig. 2). More detailed fine structure of these peaks is shown in Fig. 5(c), where the peaks from a to e are further resolved and labeled using subscripts. The second bulk and surface plasmon peaks (I and II), which were discussed above, are also presented. To interpret the interband transition peaks, the energy levels with high-density electronic states were examined from a calculated partial density-of-states (pDOS), and the possible interband electronic transitions from valence band to conduction band were investigated within the selection rules [51]. For easy comparison, the peaks predicted from the possible interband transitions are indicated as labeled from a to g in Fig. 5(c). The energies of the majority of these interband electronic transition are identified from experimental low-loss EELS spectra as well as from calculated imaginary part of the dielectric function and from peaks of $Im[-1/\varepsilon]$. The results are tabulated, and the assigned interband electronic transitions are summarized, in Table II.

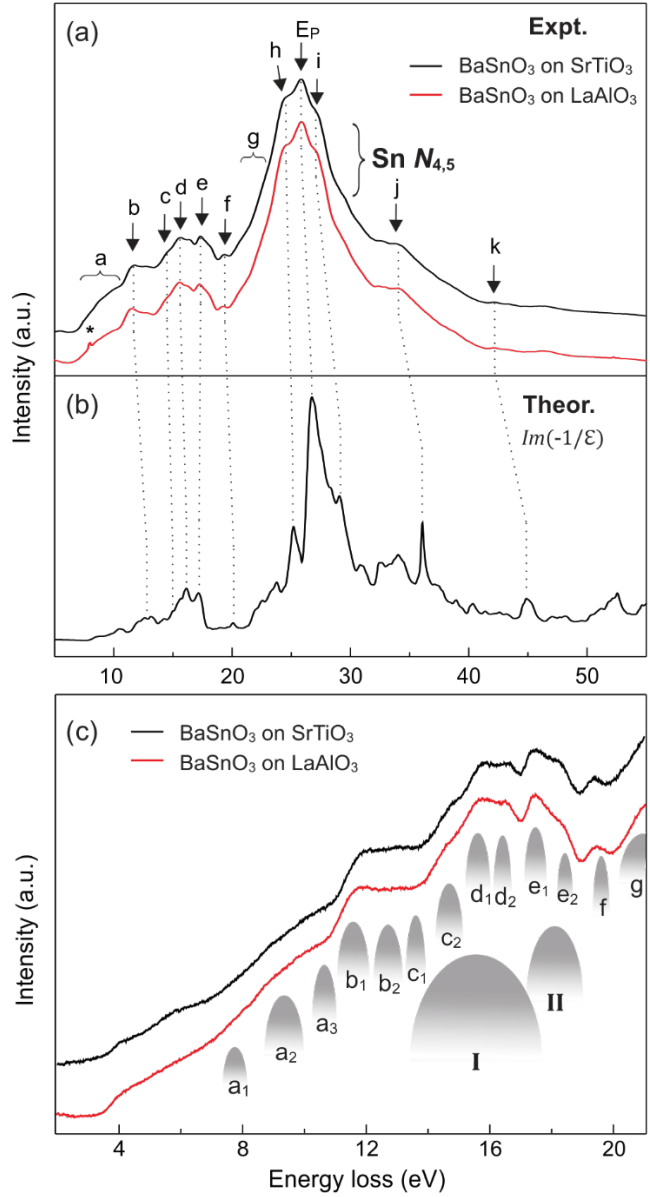

FIG. 5. Comparison of experimental low-loss EELS spectra from the two BaSnO$_3$ films (a), and calculated $Im[-1/\varepsilon]$ (b). The peaks from interband electronic transitions are labeled from a to k. The bulk plasmon peak, E$_P$, and the Sn $N_{4,5}$ edge are also indicated. The artifact peak from the ZLP is marked by an asterisk. (c) High-energy-resolution low-loss EELS spectra from the same BaSnO$_3$ films. Possible interband transitions, predicted from theory, are indicated with the labels. The peaks I and II represent predicted second bulk and surface plasmon peak, respectively.

TABLE II. Peak positions (in eV) from plasmon excitations and interband electronic transitions. The peaks are labeled as in Figs. 2 and 5 and compared to the predicted transitions from the calculated $\varepsilon$, $Im[-1/\varepsilon]$, and pDOS with assignments.

| Group | Peaks | Experiment | Theory | | | Assignment | |
|---|---|---|---|---|---|---|---|
| Plasmon | $E_P$ | 26.1 | 26.6 | | | 1st Bulk plasmon | |
| | I | - | 15.2 | | | 2nd Bulk plasmon | |
| | II | - | 18.8 | | | Surface plasmon | |
| | | | from $\varepsilon_2$ | from $Im[-1/\varepsilon]$ | from pDOS | Initial state | Final state |
| Interband | $a_1$ | - | 8.3 | 8.4 | 7.8, 8.5 | O 2$p$ | Sn 6$s$ |
| | $a_2$ | - | 9.5, 9.8 | 9.6 | 9.3, 10.1 | O 2$p$ | Sn 6$s$ |
| | $a_3$ | - | 10.4 | 10.3 | 10.2, 10.9 | O 2$p$ | Ba 5$d$ |
| | $b_1$ | ~12 | 11.2, 12.0 | 11.9, 12.4 | 11.4, 12.1, 12.7 | O 2$p$ | Sn 6$s$ |
| | | | | | 11.2, 11.7, 12.0 | O 2$p$ | Ba 5$d$ |
| | | | | | 11.4, 12.1 | Sn 4$p$ | Sn 6$s$ |
| | $b_2$ | ~13 | 12.7 | 12.9 | 12.7 | O 2$p$ | Sn 6$s$ |
| | | | | | 12.8 | O 2$p$ | Ba 5$d$ |
| | $c_1$ | ~14 | 13.6 | 13.7, 14.0 | 13.7 | O 2$p$ | Ba 5$d$ |
| | | | | | 13.6 | Sn 4$d$, Ba 4$d$ | Ba 4$f$ |
| | | | | | 13.7 | Sn 4$p$ | Ba 5$d$ |
| | $c_2$ | ~15 | 14.4, 14.8 | ~15 | 14.7 | O 2$p$ | Sn 6$s$ |
| | | | | | 14.8 | O 2$p$ | Ba 5$d$ |
| | | | | | 14.4 | Sn 4$d$, Ba 4$d$ | Ba 4$f$ |
| | | | | | 14.8 | Sn 4$p$ | Ba 5$d$ |
| | $d_1$ | 15.8 | 15.5 | 15.9 | 15.6 | Sn 4$d$, Ba 4$d$ | Ba 4$f$ |
| | $d_2$ | 16.6 | 16.5 | - | 16.4 | Sn 4$d$, Ba 4$d$ | Ba 4$f$ |
| | $e_1$ | 17.5 | 17.1 | 17.0 | 17.1, 17.8 | Sn 4$d$, Ba 4$d$ | Ba 4$f$ |
| | | | | | 17.2 | Ba 5$p$ | Sn 6$s$ |
| | $e_2$ | 18.2 | 18.5 | - | 18.5 | Sn 5$s$ | Sn 5$p$ |
| | f | 19.4 | 19.6 | 19.9 | 19.5 | O 2$p$ | Sn 5$d$ |
| | | | | | 19.4 | Sn 5$s$ | Sn 5$p$ |
| | | | | | 19.5 | Ba 5$p$ | Ba 5$d$ |
| | g | ~21 | 20.5 | - | 20.3 | O 2$p$ | Sn 5$d$ |
| | | | | | 20.6 | Ba 5$p$ | Ba 5$d$ |
| | | | | | 20.5 | Ba 5$p$ | Sn 6$s$ |
| | h | 24.7 | 25.4 | 24.9 | - | - | - |
| | i | 27.3 | 28.6 | 28.8 | - | - | - |
| | j | 34.3 | 35.7 | 35.8 | - | - | - |
| | k | 42.4 | 44.5 | 44.6 | - | - | - |

**D. Core-loss EELS analysis**

High-energy-resolution O $K$ edge EELS spectra were measured from BaSnO$_3$ films and compared with the O $K$ edge simulation result generated using the WIEN2K code, where a double differential cross section for core-level electronic transitions is calculated from the already calculated DOS; $\frac{\partial^2 \sigma}{\partial E \partial \Omega} \propto \sum_{i,f} |M_{i,f}|^2$ DOS (E), where $\Omega$ is the scattering solid angle and $M_{i,f}$ is the transition matrix element [52,53]. This double differential cross section is then integrated using experimental STEM probe convergence and EELS collection angles [25,54]. An energy spreading of the incident electron beam was implemented by convolving the simulation result with a Gaussian function with the FWHM of the ZLP. Next, the natural energy broadening, which arises from the lifetime of the electrons in the initial and final states of excitation, was taken into account [24,55,56]. The energy level width of the initial O 1$s$ state, $\Gamma_i$, is as small as 0.153 eV. In contrast, the energy level width of the final state, $\Gamma_f$, is considerable and varies with energy relative to the conduction band minimum (or onset energy of core-loss edge). The $\Gamma_f$ for O 2$p$ increases from 0 to 6 eV with increasing the energy from 0 to 40 eV above the onset energy [55]. The energy broadening due to the lifetime effect (for both initial and final states) can be represented by a Lorentzian function [44]. Thus, the simulated O $K$ edge was further convolved with Lorentzian functions with the FWHM of the $\Gamma_i$ and $\Gamma_f$, consecutively, to implement the natural energy broadening. The resulting O $K$ edge is compared to the experimental O $K$ edge EELS spectra obtained from BaSnO$_3$ films in Fig. 6. The match is good, further proving reliability of this analysis. The remaining discrepancies can be attributed to the fact that the O $K$ edge sits on the tail of the Sn $M_{4,5}$ edge and defects present in the actual samples are not accounted for in the calculations. No significant difference was observed between BaSnO$_3$ films on SrTiO$_3$ and LaAlO$_3$ substrates. The peak positions from experimental O $K$ edge EELS spectra are in good agreement with the simulation result (Table III). This comparison allows accurate determination of the onset of the EELS O $K$ edge, which represents the minimum of the conduction band, $E_C$, at 528.9 eV.

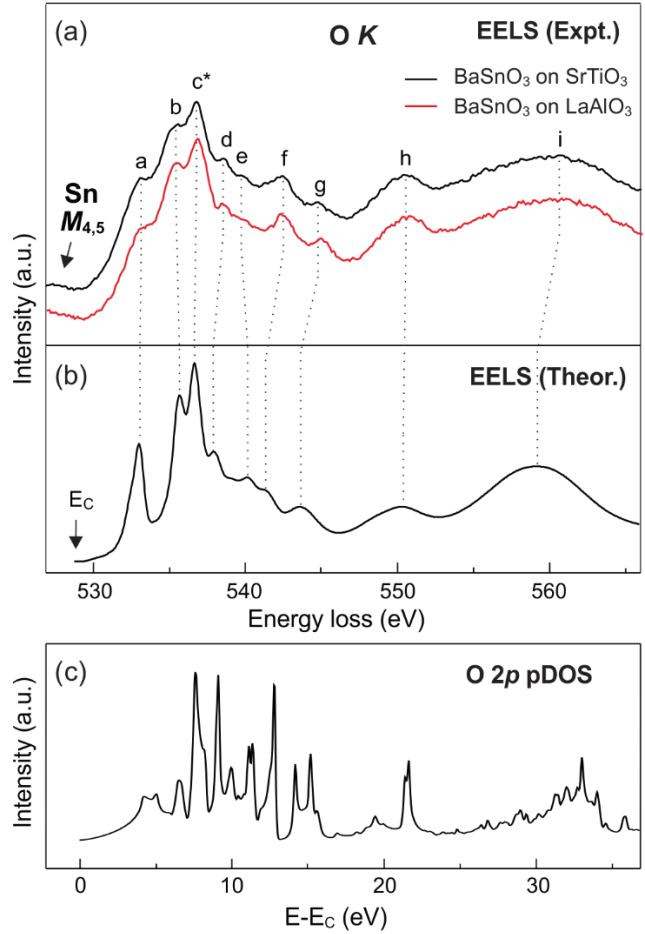

FIG. 6. Comparison of experimental EELS O $K$ edges from BaSnO$_3$ films (a) and simulated O $K$ edge (b). The peak c* was used as a reference for the alignment. The onset of the EELS O $K$ edge representing the minimum of the conduction band, $E_C$, is indicated using an arrow. The identifiable peaks are labeled and compared in Table III. Note that the O $K$ edge overlaps with the tail of the Sn $M_{4,5}$ edge. (c) The calculated O $2p$ pDOS of the conduction band in BaSnO$_3$ that was used to calculate the spectrum in (b).

TABLE III. Comparison of peak positions from the experimental and the simulated O $K$ edge EELS shown in Fig. 6. The peak c* is used for alignment.

|    | Expt. (eV) | Theor. (eV) |
|----|------------|-------------|
| a  | 533.3      | 533.2       |
| b  | 535.5      | 535.8       |
| c* | 536.8      | 536.8       |
| d  | 538.6      | 538.2       |
| e  | 539.9      | 540.4       |
| f  | 542.5      | 541.5       |
| g  | 545.0      | 543.8       |
| h  | 550.7      | 550.5       |
| i  | 560.6      | 559.4       |

The Sn $M_{4,5}$ edge and Ba $N_{4,5}$ and $M_{4,5}$ edges were also measured. The results are shown in Fig. 7. The Ba $N_{4,5}$ edge has a delayed maximum with detectable fine structure generated by the excitation of Ba 4$d$ electrons (Fig. 7(a)). The Sn $M_{4,5}$ edge, which is attributed to the excitation of Sn 3$d$ electrons, is mostly positioned just before the O $K$ edge, as shown in Fig. 7(b). The overall peak shape follows a delayed maximum shape and is similar to that from $SnO_2$ [57], indicating the presence of Sn in the 4+ oxidation state. Due to a high density of unfilled Ba 4$f$ orbitals above the Fermi energy, the Ba $M_{4,5}$ edge appears as two strong white lines: $M_5$ (labeled as n) and $M_4$ (labeled as o), separated by 15 eV. In a simple single-electron excitation description, the two peaks are explained via spin-orbit splitting [58]. The $3d^{5/2}$ ($M_5$) and $3d^{3/2}$ ($M_4$) initial states are split due to the spin-orbit interaction, with a 6:4 ratio of degeneracy of the states [24,51]. However, the ratio of integral intensity of $M_5$ and $M_4$ in experimental EELS was observed to be 0.81 for the $BaSnO_3$ film on $LaAlO_3$ and 0.83 for the $BaSnO_3$ film on $SrTiO_3$. These values deviate considerably from the 1.5 of the 6:4 ratio of degeneracy. Also, an additional pre-peak (labeled as m) is observed ahead of the $M_5$ peak. This discrepancy is primarily due to the electron-hole interaction in the final state, which is not considered in the simple single-electron excitation interpretation. The atomic multiplet theory effectively incorporates such a multi-electron excitation effect, where the electron-electron interaction, $H_{ee}$, and the spin-orbit interaction, $H_{so}$, are added into the Hamiltonian of the single-electron atomic model, $H_{1s}$, i.e. $H = H_{1s} + H_{ee} + H_{so}$ [51,58]. When the $H_{so}$ is negligible compared to the $H_{ee}$, the electronic states of an atom can be described using the LS coupling scheme, in which electronic states are determined by a given atomic configuration. By employing the LS coupling scheme, we can derive the initial and final electronic states and the available electronic transitions. Three available transitions for the excitations of Ba 4$d$ electrons to Ba 4$f$ orbitals can be predicted, which effectively describe the observations of the peaks, m, n, and o in the Ba $M_{4,5}$ edge as well as the low value of the $M_5/M_4$ ratio [58].

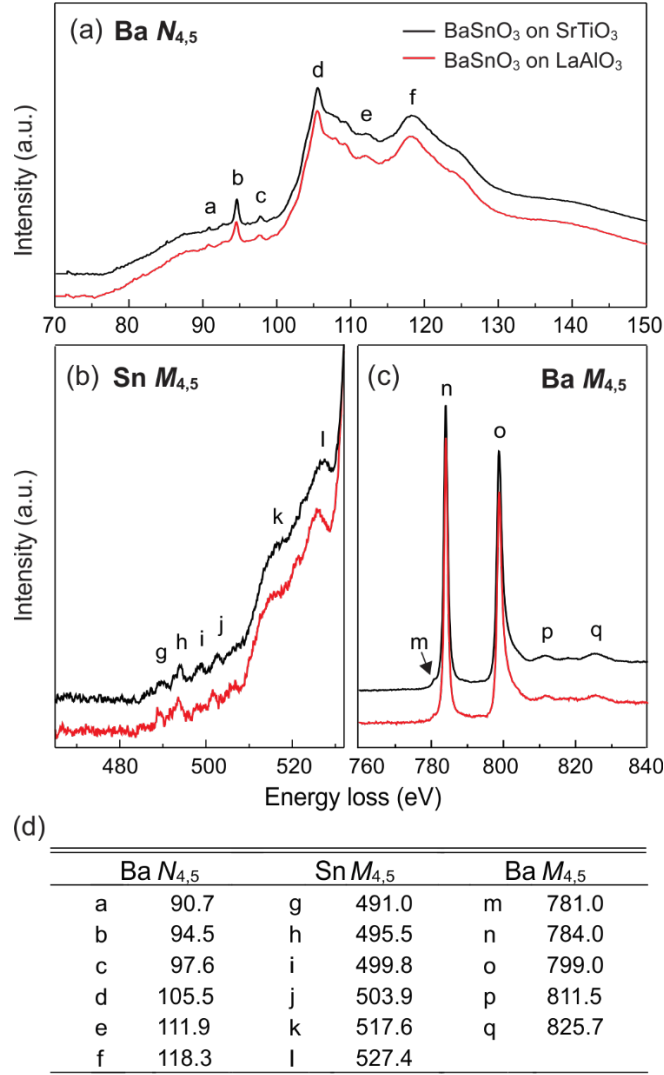

FIG. 7. Measured core-loss EELS spectra from BaSnO$_3$ films: (a) Ba $N_{4,5}$ edge, (b) Sn $M_{4,5}$ edge, and (c) Ba $M_{4,5}$ edge. (d) List of experimental peak positions of Ba $N_{4,5}$, Ba $M_{4,5}$, and Sn $M_{4,5}$ EELS edges fine structure shown in panels (a-c).

## IV. CONCLUSION

The electronic structure of epitaxial BaSnO$_3$ films grown on SrTiO$_3$(001) and LaAlO$_3$(001) substrates were investigated using experimental high-energy-resolution EELS in STEM, and *ab initio* calculations. The indirect band gap energy was measured from the low-loss EELS to be 3.0 ± 0.1 eV for the MBE-grown BaSnO$_3$ film on SrTiO$_3$ and 3.1 $\pm$ 0.2 eV for the high pressure oxygen sputtered BaSnO$_3$ film on LaAlO$_3$. Experimental low-loss EELS spectra were compared with the calculated *Im*[-1/$\varepsilon$] function, which is directly proportional to the electron energy-loss function, and the observed peaks due to plasmon excitations and interband transitions were analyzed. The experimental bulk plasmon peak was observed at 26.1 eV while the theoretical value was predicted to be at 26.6 eV. The expected small discrepancy between the experimental and calculated bulk plasmon energy was explained through plasmon damping, which was not properly taken into account in "phonon-free" and "defect-free" *ab initio* calculations. The interband electronic transition peaks were clearly observed and their positions were identified in low-loss EELS. The results were compared with predictions based on the calculated pDOS, where the observed peaks were assigned to distinct interband transitions from the valence band to conduction band.

The core-level electron excitations were also examined using the core-loss EELS edges. O *K*, Ba *N*$_{4,5}$, Sn *M*$_{4,5}$, and Ba *M*$_{4,5}$ edges were measured using high-energy-resolution EELS and their fine structures were analyzed. For comparison, a simpler O *K* edge, resulting from the excitation of O 1*s* electrons to the empty DOS above the Fermi energy, was simulated from calculated O 2*p* pDOS. When the instrumental and the natural energy broadenings were implemented into this simulation, the resulting theoretical O *K* edge was in very good agreement with the experimental O *K* edges, further conforming reliability of this analysis. The number of peaks and their relative intensities in Ba *M*$_{4,5}$ edge fine structure, which deviate from a simple spin-orbit interaction model with a 6:4 ratio of degeneracy, were explained by more rigorous atomic multiplet theory. Importantly, this work can be used as a starting point to explore the local electronic structure changes in BaSnO$_3$ films by structural defects, including dislocations, vacancies, interfaces, and impurity doping.

**ACKNOWLEDGMENTS:**

This work was supported in part by the NSF MRSEC under award number DMR-1420013, also in part by NSF DMR-1410888, NSF EAR-134866 and EAR-1319361, by Grant-in-Aid program of the University of Minnesota, and by the Defense Threat Reduction Agency, Basic Research Award #HDTRA1-14-1-0042 to the University of Minnesota. Computational resources were partly provided by Blue Waters sustained-petascale computing project, which is supported by the NSF under awards OCI-0725070 and ACI-1238993 and the state of Illinois. STEM analysis was performed in the Characterization Facility of the University of Minnesota, which receives partial support from the NSF through the MRSEC. H.Y. acknowledges a fellowship from the Samsung Scholarship Foundation, Republic of Korea.

[8] S. A. Chambers, T. C. Kaspar, A. Prakash, G. Haugstad, and B. Jalan, Band alignment at epitaxial BaSnO$_3$/SrTiO$_3$(001) and BaSnO$_3$/LaAlO$_3$(001) heterojunctions, Appl. Phys. Lett. **108**, 152104 (2016).

[9] A. Prakash, J. Dewey, H. Yun, J. S. Jeong, K. A. Mkhoyan, and B. Jalan, Hybrid molecular beam epitaxy for the growth of stoichiometric BaSnO$_3$, J. Vac. Sci. Technol. A **33**, 060608 (2015).

[10] Z. Lebens-Higgins, D. O. Scanlon, H. Paik, S. Sallis, Y. Nie, M. Uchida, N. F. Quackenbush, M. J. Wahila, G. E. Sterbinsky, D. A. Arena, J. C. Woicik, D. G. Schlom, and L. F. J. Piper, Direct observation of electrostatically driven band gap renormalization in a degenerate perovskite transparent conducting oxide, Phys. Rev. Lett. **116**, 027602 (2016).

[11] U. S. Alaan, P. Shafer, A. T. Diaye, E. Arenholz, and Y. Suzuki, Gd-doped BaSnO$_3$: A transparent conducting oxide with localized magnetic moments, Appl. Phys. Lett. **108**, 042106 (2016).

[12] J. Shiogai, K. Nishihara, K. Sato, and A. Tsukazaki, Improvement of electron mobility in La:BaSnO$_3$ thin films by insertion of an atomically flat insulating (Sr,Ba)SnO$_3$ buffer layer, AIP Advances **6**, 065305 (2016).

[13] S. Yu, D. Yoon, and J. Son, Enhancing electron mobility in La-doped BaSnO$_3$ thin films by thermal strain to annihilate extended defects, Appl. Phys. Lett. **108**, 262101 (2016).

[14] H. Mizoguchi, H. W. Eng, and P. M. Woodward, Probing the electronic structures of ternary perovskite and pyrochlore oxides containing Sn$^{4+}$ or Sb$^{5+}$, Inorg. Chem. **43**, 1667 (2004).

[15] E. Moreira, J. M. Henriques, D. L. Azevedo, E. W. S. Caetano, V. N. Freire, U. L. Fulco, and E. L. Albuquerque, Structural and optoelectronic properties, and infrared spectrum of cubic BaSnO$_3$ from first principles calculations, J. Appl. Phys. **112**, 043703 (2012).

[16] A. Slassi, Ab initio study of a cubic perovskite: Structural, electronic, optical, and electrical properties of native, lanthanum- and antimony-doped barium tin oxide, Mater. Sci. Semicond. Process. **32**, 100 (2015).

[17] S. Soleimanpour and F. Kanjouri, First principle study of electronic and optical properties of the cubic perovskite BaSnO$_3$, Physica B: Condens. Matter. **432**, 16 (2014).

[18] S. James Allen, S. Raghavan, T. Schumann, K. M. Law, and S. Stemmer, Conduction band edge effective mass of La-doped BaSnO$_3$, Appl. Phys. Lett. **108**, 252107 (2016).

[19] D. J. Singh, Q. Xu, and K. P. Ong, Strain effects on the band gap and optical properties of perovskite SrSnO$_3$ and BaSnO$_3$, Appl. Phys. Lett. **104**, 011910 (2014).

[20] T. Schumann, S. Raghavan, K. Ahadi, H. Kim, and S. Stemmer, Structure and optical band gaps of (Ba,Sr)SnO$_3$ films grown by molecular beam epitaxy, J. Vac. Sci. Technol. A **34**, 050601 (2016).
20